\documentclass[aps, twocolumn, letterpaper, 10pt]{revtex4-1}

\usepackage{graphicx}
\usepackage{amssymb, amsmath,mathtools}
\usepackage{bbm}

\newcommand{\Nc}{N_{\mathrm{c}}}
\DeclareMathOperator{\tr}{tr}

\graphicspath{{./}{./plots/}}

\begin{document}

\title{Excitonic instability of three-dimensional gapless semiconductors: Large-$N$ theory}

\author{Lukas Janssen}
\author{Igor F. Herbut}
\affiliation{Department of Physics, Simon Fraser University, Burnaby, British
Columbia, Canada V5A 1S6}

\begin{abstract}
Three-dimensional gapless semiconductors with quadratic band touching,
such as HgTe, $\alpha$-Sn, or Pr$_2$Ir$_2$O$_7$ are believed to display a 
non-Fermi-liquid ground state due to long-range electron-electron interaction. 
We argue that this state is inherently unstable towards spontaneous formation of
a (topological) excitonic insulator. The instability can be parameterized by a 
critical fermion number $\Nc$. For $N<\Nc$ the rotational symmetry is 
spontaneously broken, the system develops a gap in the spectrum, and features a 
finite nematic order parameter. To leading order in the $1/N$ expansion and 
in the static approximation, the analogy with the problem of
dynamical mass generation in (2+1)-dimensional quantum electrodynamics yields 
$\Nc = 16/[3\pi(\pi-2)]$. Taking the important dynamical screening effects into 
account, we find that $\Nc \geq 2.6 (2)$ and therefore safely above the 
physical value of $N=1$. Some experimental consequences of the nematic ground 
state are discussed.
\end{abstract}

\maketitle

\section{Introduction}
Three-dimensional (3D) Fermi systems in which valence and conduction bands
touch quadratically at the Fermi level form the very boundary
between two classes of materials: Right at the Fermi level their
density of states vanishes and the systems can hence be understood
as limiting cases of semiconductors in which the band gap goes to zero.
Away from the Fermi level, on the other hand, the density of states increases
rapidly, and the systems may alternatively be regarded as degenerate
semimetals with the band overlap region shrunk to a single point.
Prominent examples for such three-dimensional \emph{gapless semiconductors} are
given by $\alpha$-Sn and HgTe, which feature band inversion due to spin-orbit
coupling. The quadratic band touching (QBT) point, located at the
center $\Gamma$ of the Brillouin zone, is protected by crystal
symmetry, and in the undoped systems the Fermi level is right at the
touching point~\cite{tsidilkovski1997}.
When the rotational symmetry is broken, the band degeneracy at the $\Gamma$
point can be lifted and the spectrum develops a full, anisotropic, and
topologically nontrivial gap~\cite{fu2007}.
HgTe, for instance, shows the quantum spin Hall effect in
quantum well structures~\cite{koenig2007} and becomes a 3D strong topological
insulator under  uniaxial strain~\cite{bruene2011}.
The pyrochlore iridates $R_2$Ir$_2$O$_7$ (where $R$ is a rare-earth
element) that display a metallic paramagnetic state presumably also host a 3D
QBT point at the Fermi level~\cite{witczakkrempa2014, kondo2015}. This scenario has been
employed to explain the anomalous low-temperature behavior
measured in Pr$_2$Ir$_2$O$_7$~\cite{moon2013}.

The dichotomous classification of gapless semiconductors is reflected in their
concomitant peculiar \emph{marginal} screening, which allows fundamentally new
types of many-body ground states.
Theoretical control over the situation can be exerted by employing a $1/N$
expansion, where $N$ is the number of QBT points at the Fermi level. In the
limit of large $N$, Abrikosov and Beneslavskii found a scale-invariant
semimetallic ground state with anomalous power laws---a 3D non-Fermi-liquid
phase~\cite{abrikosov1971}. Recently, their analysis has been rediscovered and
extended in the context of the pyrochlore iridates~\cite{moon2013}.

In this work, we revisit the problem of the ground state of 3D
Fermi systems with QBT by deriving and exploiting the nonperturbative
solution of the Dyson-Schwinger equations. Although also controlled by
the small parameter $1/N$, this enables one to address questions
such as spontaneous symmetry breaking and dynamical mass generation.
We demonstrate that the Abrikosov-Beneslavskii large-$N$ regime exhibits a
lower bound $\Nc$ below which the non-Fermi-liquid state becomes unstable and
the system features a symmetry-breaking phase
transition. The value for $\Nc$ is found to be \emph{above} the physical $N=1$,
rendering the 3D systems with QBT insulating and nematic at low temperatures.
While such a possibility has been suggested earlier by considering the theory in
higher spatial dimensions $d>3$~\cite{herbut2014}, so far there has not been a
definite demonstration of the instability within a fully controlled
approximation. Conceptually, our scenario is analogous to the well-known
situation in (2+1)-dimensional quantum electrodynamics (QED$_3$), which displays
chiral symmetry breaking if the number of fermions is less than a certain
critical fermion number~\cite{appelquist1988, nash1989,
maris1995, grover2014, braun2014}. The principal difference is, however, the
lack of relativistic invariance at low energies, which makes the effects of
dynamical screening both crucial and nontrivial to include. We show,
nevertheless, that the critical number of four-component fermions $\Nc$ is,
to the leading order of the $1/N$ expansion, bounded from below by $2.6(2)$, and 
thus is significantly larger than the relevant physical value of $N=1$.

\section{Semiclassical picture}
The $N$-dependent excitonic transition can be understood heuristically within a
simple semiclassical
picture. At the Fermi level, the point of QBT separates the filled valence
electron band from the unoccupied conduction band. Virtual or
thermal fluctuations can induce electron-hole
pairs, which interact via an attractive screened Coulomb potential $V(r)$.
In the simplest approach, the latter is determined by
the density of states~$\rho(\varepsilon_{\vec p})$ near the
Fermi level~\cite{ashcroft1976},
\begin{equation} \label{eq:screened-potential-DOS}
 V(r) = \int \frac{d^3 \vec p}{(2\pi)^3} \frac{ e^{i{\vec r}  \cdot  { \vec p}} }{\frac{p^2
\kappa}{4\pi e^2} +
\rho(\varepsilon_{\vec p})},
\end{equation}
where $e$ is the elementary charge and $\kappa$ is the ``background''
dielectric constant arising from transitions between bands away from the
Fermi level~\cite{halperin1968}.
When the band dispersion is quadratic, $\varepsilon_{\vec p} \propto
\pm p^2$, the density of states vanishes linearly as a function of momentum,
$\rho(\varepsilon_{\vec p}) \propto N |\vec p|$, with $N$ the number of
QBT points at the Fermi level. At large distances, the form of the
potential thus becomes $\propto 1/p$ in Fourier space
or, equivalently, $V(r) \propto 1/r^2$ in real space.
The Coulomb interaction is \emph{marginally} screened, in clear contrast to
both metallic and dielectric screening.
Whether or not an electron-hole pair can form an excitonic quasiparticle is
determined by the spectrum of the corresponding $s$-wave Schr\"{o}dinger
equation for the radial ``exciton wave function''~$\Psi_\text{exc}(r)$,
\begin{equation} \label{eq:exciton-QM}
 \left[ -\frac{1}{r^2}\frac{d}{d r}\left(r^2 \frac{d}{dr}\right) -
 \frac{\alpha}{N} \frac{1}{r^2} \right] \Psi_\text{exc}(r)
 = 2m E \Psi_\text{exc}(r),
\end{equation}
where $m$ is the effective band mass, $\alpha$ is a dimensionless constant, and
we have set $\hbar=1$.
An exciton bound state would be given by the solution of 
Eq.~\eqref{eq:exciton-QM} with energy eigenvalue $E<0$.
The quantum mechanical Hamiltonian with $1/r^2$ potential is formally invariant
under the scale transformation $r \mapsto \beta r$, $\beta > 0$.
A discrete bound-state spectrum would inevitably break the scale invariance, and one
therefore might expect the spectrum to possess scattering states only.
This naive expectation, however, is correct only for weak
interaction, when  the dimensionless parameter $\alpha/N <
1/4$~\cite{landau1977}. In this case, no exciton formation is possible, and the
3D QBT system becomes scale invariant at large distances. This
is the Abrikosov-Beneslavskii non-Fermi-liquid state~\cite{abrikosov1971, moon2013}.
For $\alpha/N > 1/4$, on the other hand, the Hamiltonian
needs to be regularized at short distances and \emph{does}
admit bound states, with energy eigenvalues $E<0$ that depend on the
regularization~\cite{essin2006}. This peculiarity of the
quantum mechanics of the attractive $1/r^2$ potential can be interpreted as
possibly the simplest realization of a quantum anomaly~\cite{camblong2001}.
In our system, screening is suppressed at short distances, which  naturally
regularizes the potential and prevents the electron-hole pair from
``collapsing''~\cite{landau1977}.
For small values of $N$, exciton formation is thus possible and in fact
favored energetically.
The critical number of fermions that separates the scale-invariant
non-Fermi-liquid phase at large $N$ from the excitonic insulating phase at small
$N$ would in this approximation be given by $\Nc = 4 \alpha$.

In what  follows we use the field-theoretical machinery to
demonstrate that the above picture is
qualitatively entirely correct. In particular, we will find that the system can
be described by a differential equation formally identical to
Eq.~\eqref{eq:exciton-QM}, with a value for $\alpha$ that leads to $\Nc>1$.

\section{Field theory for 3D QBT system}
The minimal low-energy Hamiltonian relevant for semiconductors with
diamond or zincblende crystal structure is given by the Luttinger
Hamiltonian~\cite{luttinger1956, murakami2004}
\begin{equation}
 \mathcal H_0(\vec p) = \frac{1}{2m} \left[\left(\alpha_1 + \tfrac{5}{2} \alpha_2
\right)
p^2 \mathbbm 1_{4} - 2 \alpha_2 ( \vec p \cdot \vec J )^2
\right],
\end{equation}
with $\vec J$ as the ($4 \times 4$) $j=3/2$ representation of the 
angular momentum algebra, $\alpha_{i}$~the phenomenological Luttinger parameters, and where we assumed spherical symmetry. The
spectrum then is $\varepsilon_{\vec p} = (\alpha_1 \pm 2
\alpha_2) p^2 / (2m)$, and $\mathcal H_0$ describes a QBT point if $|\alpha_1|
< 2|\alpha_2|$. For large $N$, both spherical as well as
particle-hole symmetry are in fact expected to be \emph{emergent} at low
energies~\cite{moon2013}, and for simplicity we therefore also set $\alpha_1 =
0$ in the following. 
While the assumption of spherical symmetry appears to be a valid description of 
HgTe and $\alpha$-Sn~\cite{tsidilkovski1997}, we should note that strong 
correlations in the pyrochlore iridates may induce significant anisotropies~\cite{savary2014}. For a quantitative description of the latter systems 
it may therefore be necessary to go beyond our simple rotationally and particle-hole symmetric model. This is left for future work.

The Hamiltonian can then be written as~\cite{janssen2015}
\begin{equation} \label{eq:Hamiltonian-murakami}
 \mathcal H_0(\vec p) = \sum_{a=1}^{5} d_a(\vec p) \gamma_a,
\end{equation}
where $d_a(\vec p) = p^2 \tilde d_a(\vartheta, \varphi)$ are the five real
spherical harmonics for the angular momentum of two, viz., $\tilde d_1 + i
\tilde d_2 = (\sqrt{3}/2) \sin^2(\vartheta) e^{2i\varphi}$, $\tilde d_3 + i \tilde
d_4 = (\sqrt{3}/2) \sin(2\vartheta) e^{i \varphi}$, and $\tilde d_5 = (3
\cos^2\vartheta - 1)/2$, with $\vartheta$ and $\varphi$ as the spherical angles in
momentum space.
The Hermitian $4\times 4$ Dirac matrices $\gamma_1,\dots,\gamma_5$ satisfy
the Clifford algebra $\{\gamma_a,\gamma_b\} = 2 \delta_{ab}$.
For convenience, in Eq.~\eqref{eq:Hamiltonian-murakami} we
have also set the remaining effective band mass to $m/\alpha_2 = 1$.

We are interested in the effects of the long-range Coulomb interaction, which
is mediated by a scalar field~$a$, and described by
the Euclidian bare action
\begin{equation}
  S = \int d \tau d^3 \vec x \left[
  \psi^\dagger_i \left( \partial_\tau + i a + \mathcal H_0 \right) \psi_i
  + \frac{1}{2e^2} (\nabla a)^2
  \right].
\end{equation}
$\psi_i$ and $\psi^\dagger_i$ are four-component fermion fields, and
$i=1,\dots,N$. Upon integrating out $a$ in the functional integral, the bare
$1/r$ density-density interaction is recovered.

$\mathcal H_0$ contains the complete set of \emph{five} anticommuting $4\times 4$
Dirac matrices;  there is no further matrix left that would anticommute
with all matrices present in the Hamiltonian. An isotropic mass
gap, that is usually energetically favored in systems with Dirac
fermions~\cite{ryu2009} and in two-dimensional QBT models~\cite{sun2009}, is
thus impossible in the (isotropic) 3D systems with QBT. This leaves as the
energetically next-best option the full, but \emph{anisotropic} mass gap
$\Delta \propto \langle \psi^\dagger \gamma_5 \psi \rangle$~\cite{herbut2014},
which can be understood as nematic order parameter~\cite{janssen2015}.
As a genuine many-body phenomenon, dynamical mass generation is inaccessible to
perturbation theory. Similarly to QED$_3$~\cite{appelquist1988, nash1989,
maris1995} and graphene~\cite{khveshchenko2001, gorbar2002, popovici2013}, a
possible excitonic instability can, however, be revealed by a nonperturbative
solution of the gap equation for the fermion Green's function $G(\omega,\vec
p)$,
\begin{multline} \label{eq:gap-equation}
 G^{-1}(\omega,\vec p) = i \omega + \mathcal H_0(\vec p) +
\int\frac{d\nu d^3\vec q}{(2\pi)^4}
\bigl[\Gamma(\omega,\vec p;\nu, \vec q) \\
\times G(\nu,\vec q) V(\nu-\omega,\vec q-\vec p) \bigr],
\end{multline}
involving the full vertex function $\Gamma(\omega,\vec p;\nu,\vec q)$ and
Coulomb Green's function $V(\omega,\vec p)$. The latter in turn is
screened by the fermion polarization,
\begin{multline} \label{eq:polarization}
 V^{-1}(\omega,\vec p) = \frac{p^2}{e^2} - N \int\frac{d\nu d^3\vec
q}{(2\pi)^4} \tr\bigl[
 \Gamma(\omega+\nu,\vec p+ \vec q;\nu,\vec q) \\
 \times G(\omega+\nu,\vec p+\vec q) G(\nu,\vec q)\bigr].
\end{multline}
Similarly, the vertex function fulfills an equation that involves the fermion
four-point function, and in general this leads to an infinite tower of
coupled functional integral equations---the Dyson-Schwinger equations. Within
the $1/N$ expansion, the set of equations can be solved
successively. The wave-function renormalization has the expansion
$Z(\omega,\vec p) = 1 + \mathcal O(1/N)$~\cite{appelquist1988}, and to the
leading order the fermion Green's function becomes
\begin{equation} \label{eq:fermion-propagator}
G(\omega, \vec p) = \left[ i \omega + d_a(\vec p) \gamma_a +
\Delta(\omega,\vec p) \gamma_5 \right]^{-1},
\end{equation}
where we have assumed the effective band mass to be already at its
renormalized value~\cite{note1}.
In Eq.~\eqref{eq:fermion-propagator} and hereafter we adopt the summation
convention over repeated indices.
The vertex renormalization
is also suppressed by~$1/N$, $\Gamma(\omega,\vec p;\nu,\vec q)
= 1 + \mathcal O(1/N)$~\cite{appelquist1988}.
In QED$_3$, higher-order calculations give corrections to $\Nc$ of no more than
25\%~\cite{nash1989, maris1995}, and it is therefore reasonable to expect the relevant physics
in the present case also to be captured well already by the lowest-order calculation.

We first discuss the solution in the intermediate range of momenta
$\Delta(\omega,\vec p) \ll p^2 \ll e^4$, in which the gap equation can be
linearized. This assumption will be justified \emph{a~posteriori} in the limit
$N \to \Nc$. The screened Coulomb potential in this regime is, to leading order
in $1/N$,
\begin{multline} \label{eq:polarization-linearized}
 V^{-1}(\omega,\vec p) = 4 N \int_{\nu,\vec q}
\frac{\nu(\nu+\omega) - d_a(\vec q)d_a(\vec p + \vec q)}%
{\left(\nu^2+q^4\right)\left[(\nu+\omega)^2+(\vec p + \vec q)^4\right]},
\end{multline}
where we abbreviated $\int_{\nu,\vec q} \equiv \int \frac{d\nu d^3\vec
q}{(2\pi)^4}$.
Carrying out the integration gives
\begin{equation} \label{eq:screened-potential}
 V(\omega,\vec p) = \frac{1}{N|\vec p|}
\mathcal F\!\left(\sqrt{\tfrac{|\omega|}{p^2}}\right),
\end{equation}
with the dimensionless scaling function $\mathcal F$,
\begin{equation} \label{eq:scaling-function-potential}
 \mathcal F(x) \simeq
 \begin{cases}
  \frac{16\pi}{3(\pi-2)}& \text{for } x \ll 1, \\
  4\pi x& \text{for } x \gg 1,
 \end{cases}
\end{equation}
and interpolating monotonically between
these two limits for intermediate $x$.

\section{Static approximation}
In the static approximation screened potential $V(\omega,\vec p)$ and
gap function $\Delta(\omega, \vec p)$ are approximated by their values at
$\omega = 0$. Note that in real space then $V(r) \propto 1/r^2$, as
anticipated above.
If we furthermore assume $\Delta(\vec p) \equiv \Delta(|\vec p|)$, the integral
over frequency and spherical angles in the gap equation~\eqref{eq:gap-equation}
gives
\begin{equation} \label{eq:gap-equation-static-approx}
 N \Delta(p) = \frac{4}{3\pi(\pi-2)} \int_\lambda^\Lambda dq \frac{\Delta(q)
}{\max(p,q)} ,
\end{equation}
where $\Lambda\approx e^2$ and $\lambda \approx \sqrt{\Delta(0)}$ are the UV
and IR cutoffs, respectively.
This integral equation for $\Delta(p)$ is equivalent to the
differential equation~\cite{appelquist1988}
\begin{equation} \label{eq:gap-equation-differential}
 \left[\frac{d}{d p}\left(p^2 \frac{d}{dp}\right) +
\frac{4}{3\pi(\pi-2)N}\right] \Delta(p) = 0,
\end{equation}
supplemented with the boundary conditions
\begin{equation} \label{eq:gap-equation-bc}
 \lim_{p \to \infty} \left(p \frac{d\Delta(p)}{dp} + \Delta(p)
\right) = 0
\quad \text{and} \quad
\lim_{p \to 0} \Delta(p) < \infty.
% $
\end{equation}
We note that Eq.~\eqref{eq:gap-equation-differential} is formally
identical to the exciton bound-state equation \eqref{eq:exciton-QM} for $E=0$.
The very same differential equation~\eqref{eq:gap-equation-differential}
together with the boundary conditions~\eqref{eq:gap-equation-bc} appear also in
the leading-order solution of the Dyson-Schwinger equations in
QED$_3$~\cite{appelquist1988}, quenched QED$_4$~\cite{fukuda1976}, and
graphene~\cite{khveshchenko2001, gorbar2002}. We hence simply
adopt the results for the present model. For $N>\Nc=16/[3\pi(\pi-2)]$ the unique
solution to
Eqs.~\eqref{eq:gap-equation-differential} and \eqref{eq:gap-equation-bc} is the
trivial one, $\Delta(p) \equiv 0$. For $N<\Nc$, in contrast, the equations
admit the oscillatory solution
\begin{equation} \label{eq:gap-equation-solution}
 \Delta(p) \propto \frac{1}{\sqrt{p}} \sin\left[\frac{1}{2}\sqrt{\frac{\Nc}{N}
- 1}\ln\left(\frac{p}{\sqrt{\Delta(0)}}\right)\right],
\end{equation}
valid for momenta $\Delta(0) \ll p^2 \ll e^4$, and with
\begin{equation} \label{eq:mott-gap}
 \Delta(0) \propto \exp\left(\frac{-4\pi}{\sqrt{\Nc/N - 1}}
 + \mathcal O\left((\Nc/N - 1)^0\right)
\right).
\end{equation}
The higher-order terms can be estimated from the numerical solution of the gap
equation before linearization~\cite{appelquist1988} as
$\mathcal O\left((\Nc/N - 1)^0\right) = C + \mathcal O\left(\Nc/N - 1\right)$
with $C \approx 3.7$. The transition as a function of $N$ displays an \emph{essential
singularity}, similar to the Kosterlitz-Thouless transition, and the mass gap
is exponentially suppressed in the vicinity of the transition point.
This justifies the linearization of the gap equation in order
to calculate~$\Nc$.

\section{Dynamical screening}
Equations~\eqref{eq:screened-potential} and \eqref{eq:scaling-function-potential} imply
that screening is not as efficient at higher frequencies $|\omega| \gg p^2$.
As compared to the static approximation, one therefore expects the value of
$\Nc$ to \emph{increase} when dynamical screening effects are
taken into account. We demonstrate next that this is indeed the
case. The linearized gap equation that includes the full frequency dependence
of the screened potential is given by
\begin{equation} \label{eq:Fredholm-1}
 N \Delta(\omega,\vec p) = \int_{\nu,\vec q}
 \frac{\mathcal F\!\left(\sqrt{\frac{|\omega-\nu|}{(\vec p - \vec
q)^2}}\right)}{|\vec p - \vec q|(\nu^2 + q^4)}
 \Delta(\nu,\vec q),
\end{equation}
and it defines a Fredholm eigenvalue equation in the space of
integrable functions with appropriate boundary conditions. The
corresponding integral kernel can be symmetrized by rescaling the
gap function $\Delta(\omega,\vec p)/\sqrt{\omega^2+p^4} \mapsto
\Delta(\omega,\vec p)$:
\begin{equation}
 N \Delta(\omega,\vec p) = \int_{\nu,\vec q} k(\omega,\vec p;\nu,\vec q)
\Delta(\nu,\vec q),
\end{equation}
with $k(\omega,\vec p;\nu,\vec q) =
\mathcal F(   \sqrt{ \frac{|\omega-\nu|}{(\vec p - \vec q)^2}  }  ) /
[|\vec p - \vec q| \sqrt{(\nu^2 + q^4) (\omega^2 + p^4)}]$,
and the spectrum of the integral operator defined by $k$ is
thus real. Each of its eigenvectors represents a nontrivial solution to the
gap equation, and  the critical fermion number $\Nc$ is hence given by the
operator's \emph{largest eigenvalue} $\mu_\text{max}$.

\begin{figure}[t]
\includegraphics[width=.48\textwidth]{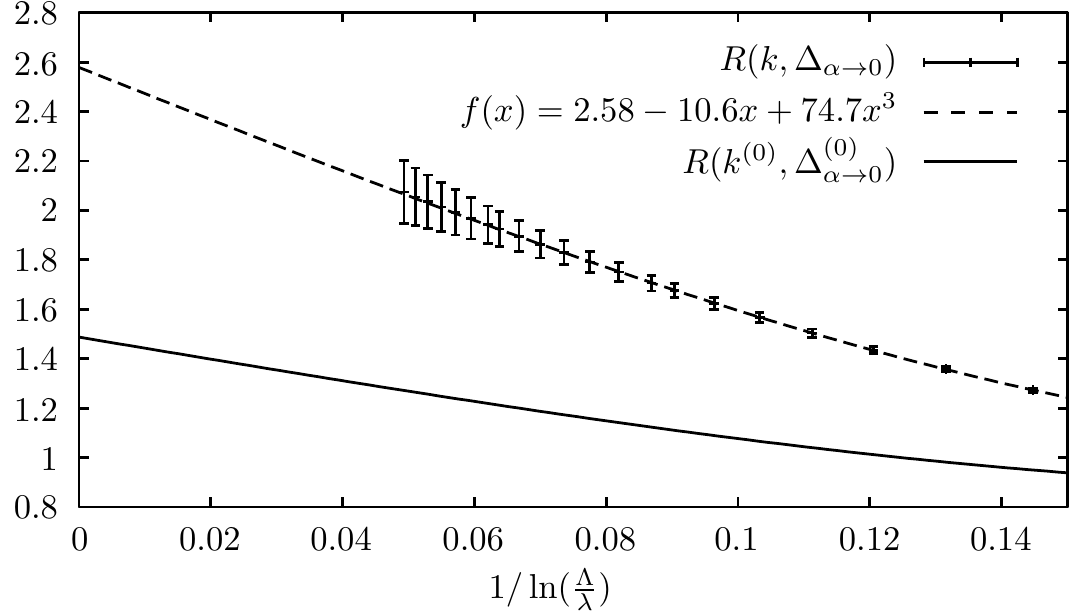}
\caption{Rayleigh quotient $R$ for $\alpha \to 0$ as a function of the cutoff
ratio $\Lambda/\lambda$ in static approximation
(solid line) and when dynamical screening effects are taken into account (data
points). For the latter, the polynomial fit $f(x) = f_0 + f_1 x + f_3 x^3$
(dashed line) yields $\Nc \geq f_0 = 2.6(2)$.}
 \label{fig:rayleigh}
\end{figure}

A lower bound on the value of $\Nc$ can be found by recognizing that the
\emph{Rayleigh quotient}
\begin{align*}
R(k,\Delta) = \frac{\langle \Delta| k | \Delta\rangle}{\langle
\Delta | \Delta\rangle}
\equiv
\frac{\int_{\omega,\vec p, \nu, \vec q} \Delta(\omega,\vec p) k(\omega,\vec
p;\nu, \vec q) \Delta(\nu,\vec q)}%
{\int_{\omega,\vec p} \Delta(\omega,\vec p) \Delta(\omega,\vec p)}
\end{align*}
is bounded from above by $R(k,\Delta) \leq \mu_\text{max} = \Nc$. This allows a
variational approach in which $R(k,\Delta)$ is maximized within a
suitable set of test functions. As proof of concept, we employ
this method first for the static approximation, where
the corresponding integral kernel is given by $k^{(0)}(p,q) =
4/[3\pi(\pi-2)\max(p,q)]$. In this case, we can construct an optimal set of
variational functions from the known solution \eqref{eq:gap-equation-solution}.
We choose $\Delta^{(0)}_{\alpha}(p) = \sin[\alpha
\log(\frac{p}{\lambda})]/\sqrt{p}$, with variational parameter $\alpha$ and IR
cutoff $\lambda$. For $\Lambda/\lambda \gg 1$ the
Rayleigh quotient is maximized when $\alpha \to 0$, and it has
the asymptotic form
\begin{equation} \label{eq:rayleigh-static}
 R(k^{(0)},\Delta^{(0)}_{\alpha\to0})
\simeq \frac{16}{3\pi(\pi-2)} \left(1 - \frac{3}{\ln(\frac{\Lambda}{\lambda})}
+ \frac{24}{\ln^3(\frac{\Lambda}{\lambda})}\right),
\end{equation}
depicted in Fig.~\ref{fig:rayleigh}.
In the limit $\Lambda/\lambda \to \infty$  the variational
method approaches the correct static-approximation result $\Nc =
16/[3\pi(\pi-2)]$, consistent with the fact that our ansatz contains
the solution~\eqref{eq:gap-equation-solution}.

In order to take the dynamical screening effects into account, we use the
related ansatz $\Delta_\alpha(\omega,\vec p)
= \Delta_{\alpha}^{(0)}(|\vec p|)/\sqrt{\omega^2 + p^4}$,
chosen such that the static solution would be recovered if dynamical
screening were again neglected in the kernel.
To obtain a lower bound on $\Nc$, we may furthermore approximate
the scaling function involved in $k(\omega,\vec p;\nu,\vec q)$ by
its lower bound $\mathcal F(x) \geq \mathcal F_0(x) = a\sqrt{1+x^2}$, with
optimized coefficient $a\simeq 12.17$,
allowing us to carry out the spherical integrals analytically.
The result of the remaining numerical integrations over frequencies
$|\omega|,|\nu| \in(\lambda^2, \Lambda^2)$ and momenta $p,q \in
(\lambda, \Lambda)$ is displayed in Fig.~\ref{fig:rayleigh} as a function
of the cutoff ratio $\Lambda/\lambda$, for the optimized variational parameter
$\alpha \to 0$. As expected, we find the Rayleigh
quotient, and therefore $\Nc$, to be significantly larger than in the static
approximation. The data points are fitted well by a third-order polynomial
inspired from Eq.~\eqref{eq:rayleigh-static}, enabling us to extrapolate to
$\Lambda/\lambda \to \infty$. This way we obtain a lower bound for the
critical fermion number with the dynamical screening effects included:
$\Nc \geq 2.6(2)$, and thus significantly above $N=1$~\cite{note2}.

\section{Experimental implications}
The low-temperature ground state is consequently described by the mean-field
Hamiltonian $\mathcal H_\text{mf} = \mathcal H_0 + \Delta_0 \gamma_5$, with
$\Delta_0 \equiv \Delta(0,0) > 0$. The spectrum of $\mathcal H_\text{mf}$ is
fully gapped, with the minimal gap in the $p_x$-$p_y$ plane. In fact,
$\mathcal H_\text{mf}$ describes a QBT system that appears as if under ``dynamically generated'' uniaxial strain~\cite{roman1972,
fu2007, moon2013}, with the rotational symmetry being \emph{spontaneously} broken.
Strained HgTe, however, is well-known as a strong
topological insulator~\cite{fu2007, bruene2011}.
Without external strain, clean HgTe and its analogs hence become at low temperatures
strong topological \textit{Mott} insulators~\cite{raghu2008, herbut2014}.
Equation~\eqref{eq:mott-gap} allows to estimate the Mott gap:
$\Delta_0 \sim \varepsilon_* \exp(-4\pi/\sqrt{\Nc/N-1} + C) \gtrsim
\varepsilon_* / 500$ when $\Nc \geq 2.6$, with $\varepsilon_*$ the characteristic
energy scale for interaction effects. For HgTe and $\alpha$-Sn it is
$\varepsilon_* \sim \mathcal O(1\,\text{meV})$~\cite{tsidilkovski1997}, but larger for
Pr$_2$Ir$_2$O$_7$~\cite{kondo2015}. Experimentally, the transition manifests
itself, for instance, in the temperature dependence of the Hall coefficient
$R_\mathrm H$, which turns from polynomial form $R_\mathrm H \propto
T^{-3/2}$ above $T_\mathrm c \sim \varepsilon_*/500 k_\mathrm B$ to an
exponential dependence $R_\mathrm H \propto \exp(\Delta_0 / k_\mathrm B T)$
below $T_\mathrm c$. Similar behavior is expected for the diagonal part of the electrical
conductivity, which should also inherit the anisotropy of the spectrum
when $T<T_\mathrm c$.

In uniform magnetic fields $\vec h$ energetics suggests that the dynamically 
induced ``strain'' aligns parallel to $\vec h$, and is described by
\begin{align}
 \mathcal H_\text{mf+h} = \mathcal H_0 + \Delta_0 \gamma_5 + h\left(J_z \cos\theta + J_z^3 \sin \theta\right),
\end{align}
Here, $J_z = i \gamma_3 \gamma_4 +
\frac{i}{2} \gamma_1 \gamma_2$, and $\theta$ controls the relative strength of
the symmetry-allowed cubic Zeeman term. 
$\mathcal H_\text{mf+h}$ leads to a rich phase diagram, previously investigated 
in the context of the QBT system under external strain~\cite{moon2013}. 
For small $\theta$ and $h > h_\mathrm{c}
\simeq \Delta_0 / \mu_\mathrm B$ two Weyl nodes with linear
dispersion along the $p_z$ axis and quadratic dispersion
perpendicular to it emerge in the spectrum---a pristine \emph{double-Weyl} semimetal
phase~\cite{huang2015}. By increasing $h$ even further, the Weyl points are
shifted away from the Fermi level and the system becomes metallic with
coexisting Fermi pockets. Across this transition the magnetic Gr\"{u}neisen
ratio~\cite{zhu2003} is conjectured to display a sign change~\cite{moon2013}.
When $\theta$ is larger, the transition for increasing $h$ is
towards a normal metal, before eventually the Weyl nodes
again emerge. The gapless non-Fermi-liquid state, by contrast, turns into a Weyl
(semi)metal already at infinitesimal magnetic field and
should not allow any transition at finite $h$ as long as $\theta$ cannot be tuned
experimentally~\cite{moon2013}.

\section{Conclusions}
We argued that the 3D gapless semiconductors with $N$ points of quadratic band
touching at the Fermi level exhibit a universal critical fermion number $\Nc$
below which the Abrikosov-Beneslavskii non-Fermi-liquid state becomes unstable.
To leading order in the $1/N$ expansion we find $\Nc = 16/[3\pi(\pi-2)]$ in the
static approximation, and even larger when dynamical screening effects are
taken into account. 
From the experience gained in the related
situation in QED$_3$~\cite{nash1989, maris1995}, we expect that our main
qualitative result $\Nc>1$ is robust upon inclusion of higher-order
corrections in $1/N$. At low temperatures, clean HgTe and its analogs consequently 
should suffer a phase
transition towards a state with spontaneously broken rotational symmetry, and with
a full but anisotropic gap in the spectrum---a topological Mott insulator phase.

We have here limited ourselves to an isotropic model, which
appears to be a valid low-energy description for
HgTe and $\alpha$-Sn, and it leads to the nematic ordering as dominant
instability.
In the pyrochlore iridates that exhibit a magnetic transition, however, quantum
fluctuations may induce significant anisotropies~\cite{savary2014}. In that
case, a mass gap that anticommutes with the effective (anisotropic)
Hamiltonian \emph{is} possible, in contrast to the
isotropic case, and hence might be favored. It can be understood as an
order parameter for the antiferromagnetic all-in/all-out
instability~\cite{savary2014}, and we therefore believe that a similar scenario
as developed in the present work might be relevant also for the transitions
found experimentally in the pyrochlores~\cite{sagayama2013}.

\acknowledgments{
The authors thank B.~Skinner for discussions and acknowledge support by
the DFG under Grants No.~JA2306/1-1 and No.~JA2306/3-1, as well as the NSERC of Canada.
Numerical integrations were performed in C using the CUBA
library~\cite{hahn2005}.}


\begin{thebibliography}{99}
\bibitem{note2}
Note that the displayed uncertainty for $\Nc$ includes only the error
from numerical integration and extrapolation, but not the systematic
error from neglecting the unknown corrections of higher order in $1/N$.
\end{thebibliography}
\end{document}